\documentclass{article}
\usepackage{amsmath}
\usepackage{amsfonts}
\usepackage{amssymb}
\usepackage{graphicx}

\input{tcilatex}
\begin{document}

\title{Towards an Informational Pragmatic Realism\thanks{%
Invited paper published in the Special Issue on \textquotedblleft Luciano
Floridi and the Philosophy of Information\textquotedblright , Mind and
Machines \textbf{24}, 37-70 (2014). }}
\author{Ariel Caticha \\
{\small Department of Physics, University at Albany-SUNY, }\\
{\small Albany, NY 12222, USA.}}
\date{}
\maketitle

\begin{abstract}
I discuss the design of the method of entropic inference as a general
framework for reasoning under conditions of uncertainty. The main
contribution of this discussion is to emphasize the pragmatic elements in
the derivation. More specifically: (1) Probability theory is designed as the
uniquely natural tool for representing states of incomplete information. (2)
An epistemic notion of information is defined in terms of its relation to
the Bayesian beliefs of ideally rational agents. (3) The method of updating
from a prior to a posterior probability distribution is designed through an
eliminative induction process that singles out the logarithmic relative
entropy as the unique tool for inference. The resulting framework includes
as special cases both MaxEnt and Bayes' rule. It therefore unifies entropic
and Bayesian methods into a single general inference scheme. I find that
similar pragmatic elements are an integral part of Putnam's \emph{internal
realism}, of Floridi's \emph{informational structural realism}, and also of
van Fraasen's \emph{empiricist structuralism}. I conclude with the
conjecture that their valuable insights can be incorporated into a single
coherent doctrine --- an \emph{informational pragmatic realism}.
\end{abstract}

\section{Introduction: an informational realism}

The awareness that the concepts of truth and reality are central to science
is quite old. In contrast, the recognition that the notion of information is
central too is still in its infancy which makes Floridi's \emph{Philosophy
of Information} particularly timely \cite{Floridi2011}. I find much to agree
with his the development of a method of levels of abstraction, the adoption
of the semantic model of theories, the valuable guidance provided by
singling out 18 open problems, including the defence of an \emph{%
informational structural realism }(see also \cite{Floridi2008}). Most
importantly, Floridi's whole argument displays a certain pragmatic attitude
that I find most appealing:\footnote{%
The Perceian slant of Floridi's pragmatism is perhaps more explicit in \cite%
{Floridi1994} and \cite{Floridi2010}.}

\begin{description}
\item[\qquad ] \textquotedblleft The mind does not wish to acquire
information for its own sake. It needs information to defend itself from
reality and survive. So information is not about representing the world: it
is rather a means to model it in such a way to make sense of it and
withstand its impact.\textquotedblright\ (\cite{Floridi2011}, p. xiv)
\end{description}

\noindent However, Floridi and I approach our subject -- information -- from
different directions. Floridi is motivated by developments in computer
science and artificial intelligence while I am mostly concerned with
physics. One might therefore expect important differences and, indeed, there
are many. For example, Floridi offers no discussion of Bayesian and entropic
methods of inference which are the main theme of this paper. Nevertheless,
rather than contradictory, I find that our approaches are sufficiently
different that, in the end, they may actually complement each other.

The connection between information and physics is natural: throughout we
will adopt the view that science consists in using information about the
world for the purpose of predicting, modeling, explaining and/or controlling
phenomena of interest to us. If this image turns out to be even remotely
accurate then we might expect that the laws of science would reflect, at
least to some extent, the methods for manipulating information. We can,
moreover, entertain the radical hypothesis that the relation between physics
and nature is considerably more indirect than usually assumed: \emph{the
laws of physics are nothing but schemes for processing information about
nature.} The evidence supporting this latter notion is already quite
considerable: most of the formal structures of thermodynamics and
statistical mechanics \cite{Jaynes1957}, and also of quantum theory (see
e.g. \cite{Caticha1998}\cite{Caticha2011}\cite{Caticha2012} and references
therein) have been derived as examples of the methods of inference discussed
later in this paper.

The basic difficulty is that inferences must be made on the basis of
information that is usually incomplete and one must learn to handle
uncertainty. Indeed, to the extent that \textquotedblleft
reality\textquotedblright\ is ultimately unknowable the incompleteness of
information turns out to be the norm rather than the exception --- which
explains why statistical descriptions such as we find in quantum mechanics
are unavoidable. The gods do play dice. And this is not surprising:\ from
the informational perspective indeterminism is natural and demands no
further explanation. Instead, what needs to be explained is why in some
special and peculiar circumstances one obtains the illusion of determinism.

The objective of the present paper is to describe the method of entropic
inference as a general framework for reasoning under conditions of
uncertainty. The main contribution of the discussion is to emphasize the
pragmatic elements in the derivation. The application of entropic methods to
derive laws of physics will not be pursued here. Instead, we will tackle
three problems. First, we describe how a scheme is designed to represent
one's state of partial knowledge\footnote{%
For our purposes we do not need to be particularly precise about the meaning
of the term `knowledge'. Note however that under a pragmatic conception of
truth there is no real difference between a `justified belief' and the more
explicit but redundant `justified true belief'.} as a web of interconnected
beliefs with no internal inconsistencies; the tools to do it are
probabilities \cite{Cox1946}\cite{Jaynes2003}. Since the source of all
difficulties is the lack of complete information two other issues must
inevitably be addressed. One concerns information itself: what, after all,
is information? The other issue is: what if we are fortunate and some
information does come our way, what do we do with it? We discuss the design
of a scheme for updating the web of beliefs. It turns out that the
instrument for updating is uniquely singled out to be entropy\footnote{%
Strictly the tool for updating is \emph{relative} \emph{entropy}. However,
as we shall later see, all entropies are relative to some prior and
therefore the qualifier relative is redundant and can be dropped. This is
somewhat analogous to the situation with energy: it is implicitly understood
that all energies are relative to some reference frame but there is no need
to constantly refer to a \emph{relative energy}.} --- from which these
entropic methods derive their name. \cite{Caticha2012} The resulting
framework includes as special cases both Jaynes' MaxEnt and Bayes' rule. It
therefore unifies entropic and Bayesian methods into a single general and
manifestly consistent inference scheme.\footnote{%
I make no attempt to provide a review of the literature on entropic
inference. The following incomplete list ref{}lects only some contributions
that are directly related to the particular approach described in this
paper: \cite{Caticha2012}, \cite{Jaynes2003}, \cite{ShoreJohnson1980}, \cite%
{Williams1980}, \cite{Skilling1988}, \cite{Rodriguez1991}, and \cite%
{CatichaGiffin2006}.}

In the final discussion I point out that closely related pragmatic elements
can be found in Putnam's \emph{internal realism} \cite{Putnam1981}, in
Floridi's \emph{informational structural realism }\cite{Floridi2008} and
also in van Fraasen's more recent \emph{empiricist structuralism} \cite%
{vanFraasen2006a}\emph{. }It might therefore be possible to pursue the
systematic development of a position -- an \emph{informational pragmatic
realism} -- that takes advantage of the valuable insights achieved in those
three doctrines.

\section{Background: the tensions within realism}

\emph{Scientific realism} means different things to different people but
most scientific realists would probably agree that (1) there exists a real
world out there that is largely independent of our thoughts, language and
point of view; and (2) and that one goal of science is to provide
descriptions of what the world is really like. van Fraasen, who is not
himself a realist, describes this form of realism as follows:

\begin{description}
\item[\qquad ] \textquotedblleft Science aims to give us, in its theories, a
literally true story of what the world is like; and acceptance of a
scientific theory involves the belief that it is true.\textquotedblright\ 
\cite{vanFraasen1980}
\end{description}

\noindent There is a tension between theses (1) and (2): Can we ever know
that our scientific descriptions are true, that is, that they match or
correspond to reality? To put it bluntly, is science at all possible?

One solution to this skeptical challenge is an empiricism that accepts
thesis (1) of an independent external reality, but denies thesis (2) that we
can actually get to know and describe it. This is a type of anti-realism. In
such a philosophy true descriptions are not likely and science, if it is to
succeed, must strive towards a more modest goal. According to van Fraasen's 
\emph{constructive empiricism}

\begin{description}
\item[\qquad ] \textquotedblleft Science aims to give us theories which are
empirically adequate; and acceptance of a theory involves as belief only
that it is empirically adequate.\textquotedblright\ \cite{vanFraasen1980}
\end{description}

The skeptical challenge also leads to a different kind of tension. On one
hand all past scientific theories have turned out to be false and therefore
it is natural to infer that our best current theories will turn out to be
false too --- this is the \emph{pessimistic meta-induction argument}. And if
all theories are ultimately false, why, then, do we even bother to do
science?

On the other hand our current theories are extremely successful. How can
this be? It seems that the only possible explanation is that our theories do
indeed capture something right about reality --- this is the \emph{%
no-miracles argument}. Indeed, as Putnam puts it:

\begin{description}
\item[\qquad ] \textquotedblleft The positive argument for realism is that
it is the only philosophy that doesn't make the success of science a
miracle.\textquotedblright\ (\cite{Putnam1975}, p.73)
\end{description}

\noindent The tension can be resolved through a compromise: our theories are
right in some respects, which explains their present success, and wrong in
others, which will explain their future failures.

In constructing new theories it would be very helpful if we could know ahead
of time which are the features in our current theories that ought to be
preserved and which discarded. The position known as \emph{structural realism%
} claims to have such knowledge. It asserts that what science somehow
manages to track faithfully is the mathematical \emph{structure} that
describes relations among entities while the intrinsic nature of these
entities remains largely unknown. Thus, structural realism (belief in a true
description of structure) is intermediate between a full-blown realism
(belief in a true description of both structure and ontology) and a
skeptical empiricism (belief only in empirical adequacy). Unfortunately, the
question of drawing a clear distinction between what constitutes intrinsic
nature and what constitutes structure is tricky; it is not even clear that
such a distinction can be drawn at all.

There are many other attempted answers --- none fully successful yet --- and
the literature on these topics is enormous. (See e.g. \cite{Papineau1996}, 
\cite{Godfrey-Smith2003}.) There are two possible directions of research
that are relevant to the theme of this paper. They arise from recognizing
the central roles played by the notion of `truth' (and its related cousin
`reality') and of `information'. Clearly, different notions of `true' and
`real' and of `information' will affect how we decide what is `true' and
what is `real' and of how `information' will help us make those decisions.

\section{Pragmatic realism}

The notion of truth adopted in scientific realism, in structural realism,
and in constructive empiricism is a `correspondence' theory of truth; true
statements are claimed to enjoy a special relation to reality: they match it
or correspond to it. Unfortunately the precise nature of such correspondence
is not clear. For, in order to say that a description matches reality, we
need to compare that description with something, and \emph{for us} this
something can only be another description. Reality-in-itself remains forever
inaccessible.

Pragmatism represents a break with this notion. There are several notions of
pragmatic truth -- and they are not free from controversy either. The
central element is a concern with the practical effectiveness of ideas as
tools for achieving our purposes, a concern with whether we ought to believe
in them or not, with their justification. I think there is much here that
can be appropriated and suitably modified to do science:

\begin{description}
\item[\qquad ] Truth is a useful idealization. True statements are those
that would be fully accepted -- that is, believed -- by ideally rational
agents under ideal epistemic conditions.
\end{description}

\noindent And, just as for other conceptual idealizations in science -- such
as inertial frames, thermal equilibrium, rigid bodies and frictionless
planes -- the fact that they are never attainable in practice does not in
the least detract from their usefulness. In fact, the notion of truth is
most useful when left a bit vague. `True' is the compliment we pay to an
idea when we are fairly certain that it is working adequately. `True' is an
idea we can trust. And there need be nothing too permanent about truth
either; if we discover that the idea is no longer working so well, then we
just withdraw the compliment and say \textquotedblleft I was wrong; I
thought it was true but it wasn't.\textquotedblright\ We do this all the
time. There is no reason why all those features of truth that are embodied
in the correspondence model should be preserved; in a truly pragmatic
account we just need to preserve what is useful.

In a pragmatic approach all pragmatic virtues, and not just empirical
adequacy, contribute to the assessment of truth:\footnote{\cite{Ellis1985}
gives a similar pragmatic position which emphasizes explanatory power.}

\begin{description}
\item[\qquad ] Science aims to give us theories that are useful for the
purposes of description, explanation, prediction, control, etc.; the
acceptance of a theory involves the conviction that the theory is indeed
useful.
\end{description}

\noindent To put it bluntly science does indeed aim to give us theories that
are true --- but the aim is for \emph{pragmatic} truth. Such truth is
objective in the sense that a statement that fails to be empirically
adequate --- and therefore not useful for predictions, control, etc. --- is
objectively false. Moreover, being objective does no mean independent of us.
The presumably true propositions derive their meaning from the theory or
conceptual framework in which they are embedded; a framework that is
designed by us for our purposes.

The notion of truth is related to that of reality. Putnam, who rejects the
label `pragmatist' for himself but is nevertheless called a `neo-pragmatist'
by others, calls his position an \emph{internal} realism because the question

\begin{description}
\item[\qquad ] \textquotedblleft ...\emph{what objects does the world
consist of?} is a question that only makes sense to ask \emph{within} a
theory or description.\textquotedblright\ \cite{Putnam1979}
\end{description}

\noindent He further explains that

\begin{description}
\item[\qquad ] \textquotedblleft In an internalist view ... signs do not
intrinsically correspond to objects, independently of how those objects are
employed and by whom. But a sign that is actually employed in a particular
way by a particular community of users can correspond to particular objects 
\emph{within the conceptual scheme of those users}. `Objects' do not exist
independently of conceptual schemes. \emph{We} cut up the world into objects
when we introduce one or another scheme of description. Since the objects 
\emph{and} the signs are alike \emph{internal} to the scheme of description,
it is possible to say what matches what.\textquotedblright\ \cite{Putnam1981}
\end{description}

\noindent Ironically, realists who favor a correspondence theory of truth
would rightfully classify internal realism and any other pragmatic realism
as forms of anti-realism.

We can be more explicit: it is we who supply the concepts of chairs and
tables but, once the concepts are in place, whether the object in front of
me is a table or a chair is a matter of objective fact about which my
(subjective) judgements can be (objectively) wrong. It is in this sense that
this doctrine is a realism --- there \emph{really} is a chair in front of me
--- but it is pragmatic realism in that the concept of chair is invented by
us for our purposes and designs. And atoms are as real as chairs. In the
middle ages the question might have been whether atoms were real but today
we have a different perspective --- the concept of atom has proved to be so
undeniably useful that we can safely assert: \emph{`real' is what atoms are}.

Naturally, a pragmatic account that denies that a detailed theory of truth
is possible or even necessary will certainly fail to satisfy someone whose
interests lie precisely in the development of such a detailed theory. But
for those of us whose interests lie in most other fields, such as science,
the pragmatic account of truth as a value judgement, as a compliment, may be
quite satisfactory --- it is all we need. Indeed, elements of pragmatism are
common in 20th century physics but this is not always sufficiently
emphasized. This is the case, for example, for both Einstein and Bohr
despite their otherwise deep and well known disagreements about the nature
of physical laws. Einstein's realism is remarkably pragmatic. He
deliberately distanced himself from a correspondence notion of truth and was
more concerned with the role of truth for the purpose of inference and
reasoning:

\begin{description}
\item[\qquad ] \textquotedblleft Truth is a quality we attribute to
propositions. When we attribute this label to a proposition we accept it for
deduction. Deduction and generally the process of reasoning is our tool to
bring cohesion to a world of perceptions. The label `true' is used in such a
way that this purpose is served best.\textquotedblright\ (quoted in \cite%
{Fine1996}, p.90)
\end{description}

\noindent Likewise, there is a close affinity between Bohr and the
pragmatism of William James (see \cite{Stapp1972}):

\begin{description}
\item[\qquad ] \textquotedblleft ... in our description of nature the
purpose is not to disclose the real essence of the phenomena but only to
track down, so far as it is possible, relations between the multiple aspects
of our experience.\textquotedblright\ (\cite{Bohr1934}, p.18)
\end{description}

\noindent and also

\begin{description}
\item[\qquad ] \textquotedblleft Owing to the very character of such
mathematical abstractions, the formalism [of quantum mechanics] does not
allow pictorial representation on accustomed lines, but aims directly at
establishing relations between observations obtained under well-defined
conditions.\textquotedblright\ (\cite{Bohr1958}, p.71)
\end{description}

\noindent a position that is clearly pragmatic and supports structuralism.

\section{The pragmatic design of probability theory}

Science requires a framework for inference on the basis of incomplete
information. Our first task is to show that the quantitative measures of 
\emph{plausibility} or \emph{degrees} \emph{of belief} that are the tools
for reasoning should be manipulated and calculated using the ordinary rules
of the calculus of probabilities --- and \emph{therefore} probabilities 
\emph{can }be interpreted as degrees of belief.

The procedure we follow differs in one remarkable way from the traditional
way of setting up physical theories. Normally one starts with the
mathematical formalism, and then one proceeds to try to figure out what the
formalism might possibly mean; one tries to append an interpretation to it.
This is a very difficult problem which has affected statistical physics ---
what is the meaning of probability and of entropy? --- and also quantum
theory --- what is the meaning of the wave function? Here we proceed in the
opposite order. First we decide what we are talking about, degrees of belief
or degrees of plausibility (we use the two expressions interchangeably) and
then we \emph{design} rules to manipulate them; we design the formalism, we
construct it to suit our purposes. The advantage of this pragmatic approach
is that the issue of meaning, of interpretation, is settled from the start.

\subsection{Rational beliefs}

The terms rational and rationality are loaded with preconceived notions. We
shall use them with a very limited and technical meaning to be explained
below. To start we emphasize that the degrees of belief discussed here are
those that would be held by an idealized \textquotedblleft
rational\textquotedblright\ agent who would not be subject to the practical
limitations under which we humans operate. Humans may hold different beliefs
and it is certainly important to figure out what those beliefs might be ---
perhaps by observing their gambling behavior --- but this is not our present
concern. Our objective is neither to assess nor to describe the subjective
beliefs of any particular individual. Instead we deal with the altogether
different but very common problem that arises when we are confused and we
want some guidance about what we are \emph{supposed }to believe. Our concern
here is not so much with beliefs as they actually are, but rather, with
beliefs as they \emph{ought} to be --- at least as they ought to be and
still deserve to be called \emph{rational}. We are concerned with an ideal
standard of rationality that we humans ought to attain at least when
discussing scientific\ matters.

The challenge is that the concept of rationality is notoriously difficult to
pin down. One thing we can say is that rational beliefs are constrained
beliefs. The essence of rationality lies precisely in the existence of some
constraints --- not everything goes. We need to identify some \emph{%
normative criteria of rationality} and the difficulty is to find criteria
that are sufficiently general to include all instances of rationally
justified belief. Here is our first criterion of rationality:

\begin{description}
\item[\qquad ] \emph{The inference framework must be based on assumptions
that have wide appeal and universal applicability.}
\end{description}

\noindent Whatever guidelines we pick they must be of general applicability
--- otherwise they would fail when most needed, namely, when not much is
known about a problem. Different rational agents can reason about different
topics, or about the same subject but on the basis of different information,
and therefore they could hold different beliefs, but they must agree to
follow the same rules --- and thus wide appeal is necessary. What we seek
here are not the specific rules of inference that will apply to this or that
specific instance; what we seek is to identify some few features that all
instances of rational inference might have in common.

The second criterion is that

\begin{description}
\item[\qquad ] \emph{The inference framework must not be self-refuting.}
\end{description}

\noindent It may not be easy to identify criteria of rationality that are
sufficiently general and precise. Perhaps we can settle for the more
manageable goal of avoiding irrationality in those glaring cases where it is
easily recognizable. And this is the approach we take: rather than providing
a precise criterion of rationality to be carefully followed, we design a
framework with the more modest goal of avoiding some forms of irrationality
that are sufficiently obvious to command general agreement. The basic desire
is that the web of rational beliefs must avoid inconsistencies. If a
conclusion can be reached in two different ways the two ways must agree. As
we shall see this requirement turns out to be extremely restrictive.

Finally,

\begin{description}
\item[\qquad ] \emph{The inference framework must be useful in practice ---
it must allow quantitative analysis.}
\end{description}

\noindent Otherwise, why bother? Incidentally, nature might very well
transcend a description in terms of a closed set of mathematical formulas.
It is not nature that demands a mathematical description; it is the
pragmatic demand that our inference schemes --- our models --- be useful
that imposes such a description.

We conclude this brief excursion into rationality with two remarks. First,
we have adopted a technical and very limited but pragmatically useful notion
of rationality which defines it in terms of avoiding certain obvious
irrationalities. Real humans are seldom rational even in this very limited
sense. And second, whatever specific design criteria are chosen, one thing
must be clear: they are justified on purely pragmatic grounds and therefore
they are meant to be only provisional. The design criteria themselves are
not immune to change and improvement. Better rational criteria will lead to
better scientific theories which will themselves lead to improved criteria
and so on. Thus, the method of science is not independent from the contents
of science.

\subsection{Quantifying rational belief}

In order to be useful we require an inference framework that allows
quantitative reasoning.\footnote{%
The argument below follows \cite{Caticha2009}. It is an elaboration of the
pioneering work of Cox \cite{Cox1946} (see also \cite{Jaynes2003}).} The
first obvious question concerns the type of quantity that will represent the
intensity of beliefs. Discrete categorical variables are not adequate for a
theory of general applicability; we need a much more refined scheme.

Do we believe proposition $a$ more or less than proposition $b$? Are we even
justified in comparing propositions $a$ and $b$? The problem with
propositions is not that they cannot be compared but rather that the
comparison can be carried out in too many different ways. We can classify
propositions according to the degree we believe they are true, their
plausibility; or according to the degree that we desire them to be true,
their utility; or according to the degree that they happen to bear on a
particular issue at hand, their relevance. We can even compare propositions
with respect to the minimal number of bits that are required to state them,
their description length. The detailed nature of our relations to
propositions is too complex to be captured by a single real number. What we
claim is that a single real number is sufficient to measure one specific
feature, the sheer intensity of rational belief. This should not be too
controversial because it amounts to a tautology: an \textquotedblleft
intensity\textquotedblright\ is precisely the type of quantity that admits
no more qualifications than that of being more intense or less intense; it
is captured by a single real number.

However, some preconception about our subject is unavoidable; we need some
rough notion that a belief is not the same thing as a desire. But how can we
know that we have captured pure belief and not belief contaminated with some
hidden desire or something else? Strictly we can't. We hope that our
mathematical description captures a sufficiently purified notion of rational
belief, and we can claim success only to the extent --- again, a pragmatic
criterion --- that the formalism proves to be useful.

The inference framework will capture two intuitions about rational beliefs.
First, we take it to be a defining feature of the intensity of \emph{%
rational }beliefs that if $a$ is more believable than $b$, and $b$ more than 
$c$, then $a$ is more believable than $c$. Such transitive rankings can be
implemented using real numbers and therefore we are led to claim that

\begin{description}
\item[\qquad ] \emph{Degrees of rational belief (or, as we shall later call
them, probabilities) are represented by real numbers.}
\end{description}

\noindent Before we proceed further we need to establish some notation. The
following choice is standard.

\subsubsection*{Notation}

For every proposition $a$ there exists its negation not-$a$, which will be
denoted $\tilde{a}{}$. If $a$ is true, then $\tilde{a}$ is false and vice
versa.

Given any two propositions $a$ and $b$ the conjunction \textquotedblleft $a$ 
\textsc{and} $b$\textquotedblright\ is denoted $ab$ or $a\wedge b$. The
conjunction is true if and only if both $a$ and $b$ are true.

Given $a$ and $b$ the disjunction \textquotedblleft $a$ \textsc{or} $b$%
\textquotedblright\ is denoted by $a\vee b$ or (less often) by $a+b$. The
disjunction is true when either $a$ or $b$ or both are true; it is false
when both $a$ and $b$ are false.

Typically we want to quantify the degrees of belief in $a$, $a\vee b$, and $%
ab$ in the context of some background information expressed in terms of some
proposition $c$ in the same universe of discourse as $a$ and $b$. Such
propositions we will write as $a|c$, $a\vee b|c$ and $ab|c$.

The real number that represents the degree of belief in $a|b$ will initially
be denoted $[a|b]$ and later in its more standard form $p(a|b)$ and all its
variations.

Degrees of rational belief will range from the extreme value $v_{F}$\ that
represents certainty that the proposition is false (for example, for any $a$%
, $[\tilde{a}|a]=v_{F}$), to the opposite extreme $v_{T}$ that represents
certainty that the proposition is true (for example, for any $a$, $%
[a|a]=v_{T}$). The transitivity of the ranking scheme implies that there is
a single value $v_{F}$ and a single $v_{T}$.

\subsubsection*{The representation of {\protect\small OR} and 
{\protect\small AND}}

The inference framework is designed to include a second intuition concerning
rational beliefs:

\begin{description}
\item[\qquad ] \emph{In order to be rational our beliefs in }$a\vee b$\emph{%
\ and }$ab$\emph{\ must be somehow related to our separate beliefs in }$a$%
\emph{\ and in }$b$\emph{.}
\end{description}

\noindent Since the goal is to design a quantitative theory, we require that
these relations be represented by some functions $F$ and $G$,%
\begin{equation}
\lbrack a\vee b|c]=F([a|c],[b|c],[a|bc],[b|ac])  \label{OR F}
\end{equation}%
and 
\begin{equation}
\lbrack ab|c]=G([a|c],[b|c],[a|bc],[b|ac])~.  \label{AND G}
\end{equation}%
Note the \emph{qualitative} nature of this assumption: what is being
asserted is the existence of some unspecified functions $F$ and $G$ and not
their specific functional forms. The same $F$ and $G$ are meant to apply to
all propositions; what is being \emph{designed} is a single inductive scheme
of universal applicability. Note further that the arguments of $F$ and $G$
include all four possible degrees of belief in $a$ and $b$ in the context of 
$c$ and not any potentially questionable subset.\footnote{%
In contrast, \cite{Cox1946} sought a representation of \textsc{AND}, $%
[ab|c]=f([a|c],[b|ac])$, and negation, $[\tilde{a}|c]=g([a|c])$.}

The functions $F$ and $G$ provide a representation of the Boolean operations 
\textsc{or} and \textsc{and}. The requirement that $F$ and $G$ reflect the
appropriate associative and distributive properties of the Boolean \textsc{%
and} and \textsc{or} turns out to be extremely constraining. Indeed, we will
show that all allowed representations are equivalent to each other and that
they are equivalent to probability theory: the associativity of \textsc{or}
requires $F$ to be equivalent to the sum rule for probabilities and the
distributivity of \textsc{and} over \textsc{or} requires $G$ to be
equivalent to the product rule for probabilities.

Our method will be \emph{design by eliminative induction}: now that we have
identified a sufficiently broad class of theories --- quantitative theories
of universal applicability, with degrees of belief represented by real
numbers and the operations of conjunction and disjunction represented by
functions --- we can start weeding the unacceptable ones out.

\subsection{The sum rule}

Our first goal is to determine the function $F$ that represents \textsc{or}.
The space of functions of four arguments is very large. Without loss of
generality we can narrow down the field to propositions $a$ and $b$ that are
mutually exclusive in the context of some other proposition $d$. Thus, 
\begin{equation}
\lbrack a\vee b|d]=F([a|d],[b|d],v_{F},v_{F})~,
\end{equation}%
which effectively restricts $F$ to a function of only two arguments, 
\begin{equation}
\lbrack a\vee b|d]=F([a|d],[b|d])~.
\end{equation}%
The restriction to mutually exclusive propositions does not represent a loss
of generality because any two arbitrary propositions can be written as the
disjunction of three mutually exclusive ones,%
\begin{equation*}
a\vee b=[(ab)\vee (a\tilde{b})]\vee \lbrack (ab)\vee (\tilde{a}b)]=(ab)\vee
(a\tilde{b})\vee (\tilde{a}b)~.
\end{equation*}%
Therefore, the general rule for the disjunction $a\vee b$ of two arbitrary
propositions can be obtained by successive applications of the special rule
for mutually exclusive propositions.

\subsubsection{The associativity constraint}

As a minimum requirement of rationality we demand that the assignment of
degrees of belief be consistent: if a degree of belief can be computed in
two different ways the two ways must agree. How else could we claim to be
rational? All functions $F$ that fail to satisfy this constraint must be
discarded.

Consider any three statements $a$, $b$, and $c$ that are mutually exclusive
in the context of a fourth $d$. The consistency constraint that follows from
the associativity of the Boolean \textsc{or}, 
\begin{equation}
(a\vee b)\vee c=a\vee (b\vee c)~,
\end{equation}%
is remarkably constraining. It essentially determines the function $F$.
Start from%
\begin{equation}
\lbrack a\vee b\vee c|d]=F\left( [a\vee b|d],[c|d]\right) =F\left(
[a|d],[b\vee c|d]\right) ~.
\end{equation}%
Use $F$ again for $[a\vee b|d]$ and also for $[b\vee c|d]$, to get 
\begin{equation}
F\{F\left( [a|d],[b|d]\right) ,[c|d]\}=F\{[a|d],F\left( [b|d],[c|d]\right)
\}~.
\end{equation}%
If we call $[a|d]=x$, $[b|d]=y$, and $[c|d]=z$, then 
\begin{equation}
F\{F(x,y),z\}=F\{x,F(y,z)\}~.  \label{assoc constraints}
\end{equation}%
Since this must hold for arbitrary choices of the propositions $a$, $b$, $c$%
, and $d$, we conclude that \emph{in order to be of universal applicability}
the function $F$ must satisfy (\ref{assoc constraints}) for arbitrary values
of the real numbers $(x,y,z)$. Therefore the function $F$ must be
associative.

\noindent \textbf{Remark: }The requirement of universality is crucial.
Indeed, in a universe of discourse with a discrete and finite set of
propositions it is conceivable that the triples $(x,y,z)$ in (\ref{assoc
constraints}) do not form a dense set and therefore one cannot conclude that
the function $F$ must be associative for arbitrary values of $x$, $y$, and $%
z $. For each specific finite universe of discourse one could design a
tailor-made, single-purpose model of inference that could be consistent,
i.e. it would satisfy (\ref{assoc constraints}), without being equivalent to
probability theory. However, we are concerned with designing a theory of
inference of universal\emph{\ }applicability, a single scheme applicable to 
\emph{all universes of discourse} whether discrete and finite or otherwise.
And the scheme is meant to be used by \emph{all rational agents}
irrespective of their state of belief --- which need not be discrete. Thus,
a framework designed for broad applicability requires that the values of $x$
form a dense set.

\subsubsection{The general solution and its regraduation}

Equation (\ref{assoc constraints}) is a functional equation for $F$. It is
easy to see that there exist an infinite number of solutions. Indeed, by
direct substitution one can easily check that eq.(\ref{assoc constraints})
is satisfied by any function of the form 
\begin{equation}
F\left( x,y\right) =\phi ^{-1}\left( \phi \left( x\right) +\phi \left(
y\right) +\beta \right) \,,  \label{xi1}
\end{equation}%
where $\phi $ is an arbitrary invertible function and $\beta $ is an
arbitrary constant. What is not so easy to to show is this is also the \emph{%
general} solution, that is, given $\phi $ one can calculate $F$ and,
conversely, given any associative $F$ one can calculate the corresponding $%
\phi $. The proof of this result is given in \cite{Caticha2012}\cite{Cox1946}%
.

The significance of eq.(\ref{xi1}) becomes apparent once it is rewritten as 
\begin{equation}
\phi \left( F\left( x,y\right) \right) =\phi \left( x\right) +\phi \left(
y\right) +\beta \quad \text{or}\quad \phi \left( \lbrack a\vee b|d]\right)
=\phi \left( \lbrack a|d]\right) +\phi \left( \lbrack b|d]\right) +\beta .
\label{regrad1}
\end{equation}%
This last form is central to any Cox-type approach to probability theory.
Note that there was nothing particularly special about the original
representation of degrees of plausibility by the real numbers $%
[a|d],[b|d],\ldots $ Their only purpose was to provide us with a ranking, an
ordering of propositions according to how plausible they are. Since the
function $\phi (x)$ is monotonic, the same ordering can be achieved using a
new set of positive numbers, 
\begin{equation}
\xi (a|d)\overset{\text{def}}{=}\phi ([a|d])+\beta ,\quad \xi (b|d)\overset{%
\text{def}}{=}\phi ([b|d])+\beta ,...  \label{regrad2}
\end{equation}%
instead of the old. The original and the regraduated scales are equivalent
because by virtue of being invertible the function $\phi $ is monotonic and
therefore preserves the ranking of propositions. However, the regraduated
scale is much more convenient because, instead of the complicated rule (\ref%
{xi1}), the \textsc{or} operation is now represented by a much simpler rule,
eq.(\ref{regrad1}), \ 
\begin{equation}
\xi \left( a\vee b|d\right) =\xi \left( a|d\right) +\xi \left( b|d\right) ~,
\label{sumxi}
\end{equation}%
which is just a sum. Thus, the new numbers are neither more nor less correct
than the old, they are just considerably more convenient.

Perhaps one can make the logic of regraduation a little bit clearer by
considering the somewhat analogous situation of introducing the quantity
temperature as a measure of degree of \textquotedblleft
hotness\textquotedblright . Clearly any acceptable measure of
\textquotedblleft hotness\textquotedblright\ must reflect its transitivity
--- if $a$ is hotter than $b$ and $b$ is hotter than $c$ then $a$ is hotter
than $c$ --- which explains why temperatures are represented by real
numbers. But the temperature scales can be quite arbitrary. While many
temperature scales may serve equally well the purpose of ordering systems
according to their hotness, there is one choice --- the absolute or Kelvin
scale --- that turns out to be considerably more convenient because it
simplifies the mathematical formalism. Switching from an arbitrary
temperature scale to the Kelvin scale is one instance of a convenient
regraduation. (\cite{Caticha2012}, p. 60)

In the old scale, before regraduation, we had set the range of degrees of
belief from one extreme of total disbelief, $[\tilde{a}|a]=v_{F}$, to the
other extreme of total certainty, $[a|a]=v_{T}$. At this point there is not
much that we can say about the regraduated $\xi _{T}=\phi (v_{T})+\beta $
but $\xi _{F}=\phi (v_{F})+\beta $ is easy to evaluate. Setting $d=\tilde{a}%
\tilde{b}$ in eq.(\ref{sumxi}) gives 
\begin{equation}
\xi (a\vee b|\tilde{a}\tilde{b})=\xi (a|\tilde{a}\tilde{b})+\xi (b|\tilde{a}%
\tilde{b})\Rightarrow \xi _{F}=2\xi _{F}~,
\end{equation}%
and therefore 
\begin{equation}
\xi _{F}=0~.  \label{xiF}
\end{equation}

\subsubsection{The general sum rule}

As mentioned earlier the restriction to mutually exclusive propositions in
the sum rule eq.(\ref{sumxi}) can be easily lifted noting that for any two 
\emph{arbitrary} propositions $a$ and $b$ we have 
\begin{equation}
a\vee b=(ab)\vee (a\tilde{b})\vee (\tilde{a}b)=a\vee (\tilde{a}b)
\end{equation}%
Since each of the two terms on the right are mutually exclusive the sum rule
(\ref{sumxi}) applies, 
\begin{eqnarray}
\xi (a\vee b|d) &=&\xi (a|d)+\xi (\tilde{a}b|d)+[\xi (ab|d)-\xi (ab|d)] 
\notag \\
&=&\xi (a|d)+\xi (ab\vee \tilde{a}b|d)-\xi (ab|d)~,
\end{eqnarray}%
which leads to the general sum rule, 
\begin{equation}
\xi (a\vee b|d)=\xi (a|d)+\xi (b|d)-\xi (ab|d)~.  \label{sum rule}
\end{equation}

\subsection{The product rule}

Next we consider the function $G$ in eq.(\ref{AND G}) that represents 
\textsc{and}. Once the original plausibilities are regraduated by $\phi $
according to eq.(\ref{regrad2}), the new function $G$ for the plausibility
of a conjunction reads%
\begin{equation}
\xi (ab|c)=G[\xi (a|c),\xi (b|c),\xi (a|bc),\xi (b|ac)]~.
\end{equation}%
The space of functions of four arguments is very large so we first narrow it
down to just two. Then we require that the representation of \textsc{and} be
compatible with the representation of \textsc{or} that we have just
obtained. This amounts to imposing a consistency constraint that follows
from the distributive properties of the Boolean \textsc{and} and \textsc{or.}
A final trivial regraduation yields the product rule of probability theory.

The derivation proceeds in two steps. First, we separately consider special
cases where the function $G$ depends on only two arguments, then three, and
finally all four arguments. Using commutivity, $ab=ba$, the number of
possibilities can be reduced to seven: 
\begin{eqnarray*}
\xi (ab|c) &=&G^{(1)}[\xi (a|c),\xi (b|c)] \\
\xi (ab|c) &=&G^{(2)}[\xi (a|c),\xi (a|bc)] \\
\xi (ab|c) &=&G^{(3)}[\xi (a|c),\xi (b|ac)] \\
\xi (ab|c) &=&G^{(4)}[\xi (a|bc),\xi (b|ac)] \\
\xi (ab|c) &=&G^{(5)}[\xi (a|c),\xi (b|c),\xi (a|bc)] \\
\xi (ab|c) &=&G^{(6)}[\xi (a|c),\xi (a|bc),\xi (b|ac)] \\
\xi (ab|c) &=&G^{(7)}[\xi (a|c),\xi (b|c),\xi (a|bc),\xi (b|ac)]
\end{eqnarray*}%
It is rather straightforward to show \cite{Caticha2009}\cite{Caticha2012}
that the only functions $G$ that are viable candidates for a general theory
of inductive inference are equivalent to type $G^{(3)}$, 
\begin{equation}
\xi (ab|c)=G[\xi (a|c),\xi (b|ac)]~.  \label{G xi}
\end{equation}

The \textsc{and} function $G$ will be determined by requiring that it be
compatible with the regraduated \textsc{or} function $F$, which is just a
sum. Consider three statements $a$, $b$, and $c$, where the last two are
mutually exclusive, in the context of a fourth, $d$. Distributivity of 
\textsc{and} over \textsc{or}, 
\begin{equation}
a\left( b\vee c\right) =ab\vee ac~,
\end{equation}%
implies that $\xi \left( a\left( b\vee c\right) |d\right) $ can be computed
in two ways, 
\begin{equation}
\xi \left( a\left( b\vee c\right) |d\right) =\xi \left( \left( ab|d\right)
\vee \left( ac|d\right) \right) ~.
\end{equation}%
Using eq.(\ref{sumxi}) and (\ref{G xi}) leads to 
\begin{equation*}
G[\xi \left( a|d\right) ,\xi \left( b|ad\right) +\xi \left( c|ad\right) ]=%
\text{ }G[\xi \left( a|d\right) ,\xi \left( b|ad\right) ]+G[\xi \left(
a|d\right) ,\xi \left( c|ad\right) ]~,
\end{equation*}%
which we rewrite as%
\begin{equation}
G\left( u,v+w\right) =G\left( u,v\right) +G\left( u,w\right) ~,
\label{dist R}
\end{equation}%
\newline
where $\xi \left( a|d\right) =u$, $\xi \left( b|ad\right) =v$, and $\xi
\left( c|ad\right) =w$.

To solve the functional equation (\ref{dist R}) we first transform it into a
differential equation. Differentiate with respect to $v$ and $w$, 
\begin{equation}
\frac{\partial ^{2}\,G\left( u,v+w\right) }{\partial v\partial w}=0~,
\end{equation}%
and let $v+w=z$, to get 
\begin{equation}
\frac{\partial ^{2}\,G\left( u,z\right) }{\partial z^{2}}=0~,
\end{equation}%
which shows that $G$ is linear in its second argument, 
\begin{equation}
G(u,v)=A(u)v+B(u)~.
\end{equation}%
Substituting back into eq.(\ref{dist R}) gives $B(u)=0$. To determine the
function $A(u)$ we note that the degree to which we believe in $ad|d$ is
exactly the degree to which we believe in $a|d$ by itself. Therefore, 
\begin{equation}
\xi (a|d)=\xi (ad|d)=G[\xi (a|d),\xi (d|ad)]=G[\xi (a|d),\xi _{T}]~\,,
\end{equation}%
where $\xi _{T}$ denotes complete certainty. Equivalently, 
\begin{equation}
u=A(u)\xi _{T}\Rightarrow A(u)=\frac{u}{\xi _{T}}~.
\end{equation}%
Therefore,%
\begin{equation}
G\left( u,v\right) =\frac{uv}{\xi _{T}}\quad \text{or\quad }\frac{\xi \left(
ab|d\right) }{\xi _{T}}=\frac{\xi \left( a|d\right) }{\xi _{T}}\frac{\xi
\left( b|ad\right) }{\xi _{T}}~\,.
\end{equation}%
The constant $\xi _{T}$ is easily regraduated away: just normalize $\xi $ to 
$p=\xi /\xi _{T}$. The corresponding regraduation of the sum rule, eq.(\ref%
{sum rule}) is equally trivial. The degrees of belief $\xi $ range from
total disbelief $\xi _{F}=0$ to total certainty $\xi _{T}$. The
corresponding regraduated values are $p_{F}=0$ and $p_{T}=1$.

\subsection{Probabilities}

In the regraduated scale the \textsc{and} operation is represented by a
simple product rule, \ 
\begin{equation}
p\left( ab|d\right) =p\left( a|d\right) p\left( b|ad\right) ~,
\label{p rule}
\end{equation}%
\newline
and the \textsc{or} operation is represented by the sum rule, 
\begin{equation}
p\left( a\vee b|d\right) =p\left( a|d\right) +p\left( b|d\right) -p(ab|d)~.
\label{s rule}
\end{equation}%
\newline
Degrees of belief $p$ measured in this particularly convenient regraduated
scale will be called \textquotedblleft probabilities\textquotedblright .\
The degrees of belief $p$ range from total disbelief $p_{F}=0$ to total
certainty $p_{T}=1$.

\noindent To summarize:

\begin{description}
\item[\qquad ] \emph{A state of partial knowledge --- a web of
interconnected rational beliefs --- is mathematically represented by
quantities that are to be manipulated according to the rules of probability
theory. Degrees of rational belief are probabilities. }
\end{description}

\noindent Other equivalent representations are possible but less convenient;
the choice is made on purely pragmatic grounds.

\subsubsection*{On meaning, ignorance and randomness}

The product and sum rules can be used as the starting point for a theory of
probability: Quite independently of what probabilities could possibly mean,
we can develop a formalism of real numbers (measures) that are manipulated
according to eqs.(\ref{p rule}) and (\ref{s rule}). This is the approach
taken by Kolmogorov. The advantage is mathematical clarity and rigor. The
disadvantage, of course, is that in actual applications the issue of
meaning, of interpretation, turns out to be important because it affects how
and why probabilities are used. It affects how one sets up the equations and
it even affects our perception of what counts as a solution.

The advantage of the approach described above is that the issue of meaning
is clarified from the start: the theory was designed to apply to degrees of
belief. Consistency requires that these numbers be manipulated according to
the rules of probability theory. This is all we need. There is no reference
to measures of sets or large ensembles of trials or even to random
variables. This is remarkable: it means that we can apply the powerful
methods of probability theory to reasoning about problems where nothing
random is going on, and to single events for which the notion of an ensemble
is either absurd or at best highly contrived and artificial. Thus,
probability theory is \emph{the} method for consistent reasoning in
situations where the information available might be insufficient to reach
certainty: probability is \emph{the} tool for coping with uncertainty and
ignorance.

The degree-of-belief interpretation can also be applied to the probabilities
of random variables. It may, of course, happen that there is an unknown
influence that affects a system in unpredictable ways and that there is a
good reason why this influence remains unknown, namely, that it is so
complicated that the information necessary to characterize it cannot be
supplied. Such an influence we can call `random'. Thus, being random is just
one among many possible reasons why a quantity might be uncertain or unknown.

\section{What is information?}

The term `information' is used with a wide variety of different meanings
(see \emph{e.g.}, \cite{Floridi2011} \cite{Caticha2012} \cite{Jaynes2003} 
\cite{CoverThomas1991} \cite{Golan2008} \cite{Adriaans2008}). There is the
Shannon notion of information, which is meant to measure an amount of
information and is quite divorced from semantics. There is also an
algorithmic notion of information, which captures the notion of complexity
and originates in the work of Solomonov, Kolmogorov and Chaitin. Here we
develop an epistemic notion of information that is somewhat closer to the
everyday colloquial use of the term --- roughly, information is what we seek
when we ask a question.

It is not unusual to hear that systems \textquotedblleft
carry\textquotedblright\ or \textquotedblleft contain\textquotedblright\
information or that \textquotedblleft information is
physical\textquotedblright . This mode of expression can perhaps be traced
to the origins of information theory in Shannon's theory of communication.
We say that we have received information when among the vast variety of
messages that could conceivably have been generated by a distant source, we
discover which particular message was actually sent. It is thus that the
message \textquotedblleft carries\textquotedblright\ information. The
analogy with physics is immediate: the set of all possible states of a
physical system can be likened to the set of all possible messages, and the
actual state of the system corresponds to the message that was actually
sent. Thus, the system \textquotedblleft conveys\textquotedblright\ a
message: the system \textquotedblleft carries\textquotedblright\ information
about its own state. Sometimes the message might be difficult to read, but
it is there nonetheless.

This language --- information is physical\ --- useful as it has turned out
to be, does not exhaust the meaning of the word `information'. The goal of
Shannon's information theory, or better, communication theory, is to
characterize the sources of information, to measure the capacity of the
communication channels, and to learn how to control the degrading effects of
noise. It is somewhat ironic but nevertheless true that this
\textquotedblleft information\textquotedblright\ theory\ is unconcerned with
the central Bayesian issue of how messages affect the beliefs of rational
agents.

A fully Bayesian information theory demands an explicit account of the
relation between information and the beliefs of ideally rational agents. The
connection arises as follows. Implicit in the recognition that most of our
beliefs are held on the basis of incomplete information is the idea that our
beliefs would be \textquotedblleft better\textquotedblright\ if only we had
more information. Indeed, it is a presupposition of thought itself that some
beliefs are better than others --- otherwise why go through the trouble of
thinking? Therefore a theory of probability demands a theory for updating
probabilities.

The concern with `good' and `better' bears on the issue of whether
probabilities are subjective, objective, or somewhere in between. We can
argue that what makes one probability assignment better than another is that
it better reflects something \textquotedblleft objective\textquotedblright\
about the world. The adoption of better beliefs has real consequences: they
provide a better guidance about how to cope with the world, and in this
pragmatic sense, they provide a better guide to the \textquotedblleft
truth\textquotedblright . Probabilities are useful to the extent that they
incorporate some degree of objectivity.

On the other hand all probability assignments involve judgements and
therefore some subjectivity is unavoidable. The long controversy over the
objective or subjective nature of probabilities arises from accepting the
sharp dichotomy that they are either one or the other with no room for
intermediate positions. The Gordian knot is cut by simply declaring that 
\emph{the dichotomy is false}. What we have is something like a spectrum.
Some subjectivity is inevitable but objectivity is the desirable goal. The
procedure for enhancing objectivity is through appropriate updating
mechanisms that allow us to process information and incorporate its
objective features into our beliefs. Barring some pathological cases Bayes'
rule behaves precisely in this way. Indeed, as more and more data are taken
into account the original (possibly subjective) prior becomes less and less
relevant, and all rational agents become more and more convinced of the 
\emph{same} truth. This is crucial: were it not this way Bayesian reasoning
would not be deemed acceptable.

To set the stage for the discussion below assume that we have received a
message --- but the carrier of information could equally well have been
input from our senses or data from an experiment. If the message agrees with
our prior beliefs we can safely ignore it. The message is boring; it carries
no news; literally, for us it carries no information. The interesting
situation arises when the message surprises us; it is not what we expected.
A message that disagrees with our prior beliefs presents us with a problem
that demands a decision. If the source of the message is not deemed reliable
then the contents of the message can be safely ignored --- it carries no
information; it is no different from noise. On the other hand, if the source
of the message is deemed reliable then we have an opportunity to improve our
beliefs --- we ought to update our beliefs to agree with the message.
Choosing between these two options requires a decision, a judgement. The
message (or the sensation, or the data) becomes \textquotedblleft
information\textquotedblright\ precisely at that moment when as a result of
our evaluation we feel that our beliefs require revision.

We are now ready to address the question: What, after all, is `information'?
The main observation is that the result of being confronted with new
information\ is to restrict our options as to what we are honestly and
rationally allowed to believe. This, I propose, is the defining
characteristic of information.

\begin{description}
\item[\qquad ] \emph{Information, in its most general form, is whatever
affects and therefore constrains rational beliefs}.
\end{description}

\noindent Since our objective is to update from a prior distribution to a
posterior when \emph{new} information becomes available we can state that

\begin{description}
\item[\qquad ] \emph{New information is what forces a change of rational
beliefs. }\noindent
\end{description}

\noindent New information is a set of constraints on the family of
acceptable posterior distributions. Our definition captures an idea of
information that is directly related to changing our (rational) minds:
information is the driving force behind the process of learning.
Incidentally, note that although we did not find it necessary to talk about
amounts of information, whether measured in units of bits or otherwise, our
notion of information will allow precise quantitative calculations. Indeed,
constraints on the acceptable posteriors are precisely the kind of
information the method of maximum entropy (to be developed below) is
designed to handle.

An important aspect of this epistemic notion of information is that the
identification of what qualifies as information --- as opposed to mere noise
--- already involves a judgement, an evaluation; it is a matter of facts as
much as a matter of values. Furthermore, once a certain proposition has been
identified as information, the revision of beliefs acquires a moral
component; it is no longer optional: it becomes a moral imperative.

The act of updating is a type of dynamics --- the study of change. In
Newtonian dynamics the state of motion of a system is described in terms of
its momentum --- the \textquotedblleft quantity\textquotedblright\ of motion
--- while the change from one state to another is explained in terms of an
applied force. Similarly, a state of belief is described in terms of
probabilities --- a \textquotedblleft quantity\textquotedblright\ of belief
--- and the change from one state to another is due to information. Just as
a force or an impulse is that which induces a change from one state of
motion to another, so \emph{information is that which induces a change from
one state of belief to another}. Updating is a form of dynamics --- and vice
versa: in \cite{Caticha2011} quantum dynamics is derived as an updating of
probabilities subject to the appropriate information/constraints.

What about prejudices and superstitions? What about divine revelations? Do
they constitute information? Perhaps they lie outside our restriction to the
beliefs of \emph{ideally rational agents}, but to the extent that their
effects are indistinguishable from those of other sorts of information,
namely, they affect beliefs, they should qualify as information too. False
information is information too. Assessing whether the sources of such
information are reliable or not can be a difficult problem. In fact, even
ideally rational agents can be affected by false information because the
evaluation that assures them that the data was competently collected or that
the message originated from a reliable source involves an act of judgement
that is not completely infallible. Strictly, all judgements that constitute
the necessary first step of one inference process, are themselves the end
result of a previous inference process that is not immune from uncertainty.

What about limitations in our computational power? Such practical
limitations are unavoidable and they do influence our inferences. Should
they be considered information? No. Limited computational resources may
affect the numerical approximation to the value of, say, an integral, but
they do not affect the actual value of the integral. Similarly, limited
computational resources may affect the approximate imperfect reasoning of
real agents and real computers but they do not affect the reasoning of those
ideal rational agents that are the subject of our present concerns.

\section{The design of entropic inference}

Once we have decided, as a result of the confrontation of new information
with old beliefs, that our beliefs require revision the problem becomes one
of deciding how precisely this ought to be done. First we identify some
general features of the kind of belief revision that one might consider
desirable, of the kind of belief revision that one might count as rational.
Then we design a method, a systematic procedure, that implements those
features. To the extent that the method performs as desired we can claim
success. The point is not that success derives from our method having
achieved some intimate connection to the inner wheels of reality; success
just means that the method seems to be working. Whatever criteria of
rationality we choose, they are meant to be only provisional --- they are
not immune from further change and improvement.

Typically the new information will not affect our beliefs in just one
proposition --- in which case the updating would be trivial. Tensions
immediately arise because the beliefs in various propositions are not
independent; they are interconnected by demands of consistency --- the sum
and product rules we derived earlier. Therefore the new information also
affects our beliefs in all those \textquotedblleft
neighboring\textquotedblright\ propositions that are directly linked to it,
and these in turn affect their neighbors, and so on. The effect can
potentially spread over the whole network of beliefs; it is the whole web of
beliefs that must be revised.

The one obvious requirement is that the updated beliefs ought to agree with
the newly acquired information. Unfortunately, this requirement, while
necessary, is not sufficiently restrictive: we can update in many ways that
preserve both internal consistency and consistency with the new information.
Additional criteria are needed. What rules is it rational to choose?

\subsection{General criteria\ }

The rules are motivated by the same pragmatic design criteria that motivate
the design of probability theory itself --- universality, consistency, and
practical utility. But this is admittedly too vague; we must be more
specific about the precise way in which they are implemented.

\subsubsection{Universality}

The goal is to design a method for induction, for reasoning when not much is
known. In order for the method to perform its function we must impose that
it be of \emph{universal }applicability. Consider the alternative: We could
design methods that are problem-specific, and employ different induction
methods for different problems. Such a framework, unfortunately, would fail
us precisely when we need it most, namely, in those situations where the
information available is so incomplete that we do not know which method to
employ.

We can argue this point somewhat differently: It is quite conceivable that
different situations could require different problem-specific induction
methods. What we want to design here is a general-purpose method that
captures what all those problem-specific methods have in common.

\subsubsection{Parsimony}

To specify the updating we adopt a very conservative criterion that
recognizes the value of information: what has been laboriously learned in
the past is valuable and should not be disregarded unless rendered obsolete
by new information. The only aspects of one's beliefs that should be updated
are those for which new evidence has been supplied. Thus we adopt a

\begin{description}
\item[\textbf{Principle of Minimal Updating}:] \emph{Beliefs should be
updated only to the extent required by the new information.}
\end{description}

\noindent The special case of updating in the absence of new information
deserves special attention. It states that when there is no new information
an ideally rational agent should not change its mind.\footnote{%
We refer to ideally rational agents who have fully processed all information
acquired in the past. Humans do not normally behave this way; they often
change their minds by processes that are not fully conscious.} In fact, it
is difficult to imagine any notion of rationality that would allow the
possibility of changing one's mind for no apparent reason. This is important
and it is worthwhile to consider it from a different angle. Degrees of
belief, probabilities, are said to be subjective: two different individuals
might not share the same beliefs and could conceivably assign probabilities
differently. But subjectivity does not mean arbitrariness. It is not a blank
check allowing the rational agent to change its mind for no good reason.

Minimal updating offers yet another pragmatic advantage. As we shall see
below, rather than identifying what features of a distribution are singled
out for updating and then specifying the detailed nature of the update, we
will adopt design criteria that stipulate what is not to be updated. The
practical advantage of this approach is that it enhances objectivity ---
there are many ways to change something but only one way to keep it the same.

The analogy with mechanics can be pursued further: if updating is a form of
dynamics, then minimal updating is a form of inertia. Rationality and
objectivity demand a considerable amount of inertia.

\subsubsection{Independence}

The next general requirement turns out to be crucially important: without it
the very possibility of scientific theories would not be possible. The point
is that in every scientific model, whatever the topic, if it is to be useful
at all, we must assume that all relevant variables have been taken into
account and that whatever was left out --- the rest of the universe --- does
not matter. To put it another way: in order to do science we must be able to
understand parts of the universe without having to understand the universe
as a whole. Granted, it is not necessary that the understanding be complete
and exact; it must just be adequate for our purposes.

The assumption, then, is that it is possible to focus our attention on a
suitably chosen system of interest and neglect the rest of the universe
because they are \textquotedblleft sufficiently
independent\textquotedblright . Thus, in any form of science the notion of
statistical independence must play a central and privileged role. This idea
--- that some things can be neglected, that not everything matters --- is
implemented by imposing a criterion that tells us how to handle independent
systems. The requirement is quite natural: \emph{Whenever two systems are a
priori believed to be independent and we receive information about one it
should not matter if the other is included in the analysis or not.} This
amounts to requiring that independence be preserved unless information about
correlations is explicitly introduced.\footnote{%
The independence requirement is rather subtle and one must be careful about
its precise implementation. The robustness of the design is shown by
exhibiting an alternative version that takes the form of a consistency
constraint:\ \emph{Whenever systems are known to be independent it should
not matter whether the analysis treats them jointly or separately}. \cite%
{Caticha2012}\cite{CatichaGiffin2006}}

Again we emphasize:\ none of these criteria are imposed by Nature. They are
desirable for pragmatic reasons; they are imposed by design.

\subsection{Entropy as a tool for updating probabilities}

Consider a variable $x$ the value of which is uncertain; $x$ can be discrete
or continuous, in one or in several dimensions. It could, for example,
represent the possible microstates of a physical system, a point in phase
space, or an appropriate set of quantum numbers. The uncertainty about $x$
is described by a probability distribution $q(x)$. Our goal is to update
from the prior distribution $q(x)$ to a posterior distribution $P(x)$ when
new information becomes available. The information is in the form of a set
of constraints that defines a family $\{p(x)\}$ of acceptable distributions
and the question is: which distribution $P\in \{p\}$ should we select?

Our goal is to design a method that allows a systematic search for the
preferred posterior distribution. The central idea, first proposed in \cite%
{Skilling1988},\footnote{\cite{Skilling1988} deals with the more general
problem of ranking positive additive distributions which also include, \emph{%
e.g.}, intensity distributions.} is disarmingly simple: to select the
posterior first rank all candidate distributions in increasing \emph{order
of preference} and then pick the distribution that ranks the highest.
Irrespective of what it is that makes one distribution preferable over
another (we will get to that soon enough) it is clear that any ranking
according to preference must be transitive: if distribution $p_{1}$ is
preferred over distribution $p_{2}$, and $p_{2}$ is preferred over $p_{3}$,
then $p_{1}$ is preferred over $p_{3}$. Such transitive rankings are
implemented by assigning to each $p(x)$ a real number $S[p]$ in such a way
that if $p_{1}$ is preferred over $p_{2}$, then $S[p_{1}]>S[p_{2}]$. The
functional $S[p]$ will be called the entropy of $p$. The selected
distribution (one or possibly many, for there may be several equally
preferred distributions) is that which maximizes the entropy functional.

The importance of this particular approach to updating distributions cannot
be overestimated: it implies that the updating method will take the form of
a variational principle --- the method of Maximum Entropy (ME) --- involving
a certain functional --- the entropy --- that maps distributions to real
numbers and that is designed to be maximized. \emph{These features are not
imposed by Nature; they are all imposed by design. }They are dictated by the
function that the ME method is supposed to perform. (Thus, it makes no sense
to seek a generalization in which entropy is a complex number or a vector;
such a generalized entropy would just not perform the desired function.)

Next we specify the ranking scheme, that is, we choose a specific functional
form for the entropy $S[p]$. Note that \emph{the purpose of the method is to
update from priors to posteriors} so the ranking scheme must depend on the
particular prior $q$ and therefore the entropy $S$ must be a functional of
both $p$ and $q$. The entropy $S[p,q]$ describes a ranking of the
distributions $p$ \emph{relative} to the given prior $q$. $S[p,q]$ is the
entropy of $p$ \emph{relative} to $q$, and accordingly $S[p,q]$ is commonly
called a \emph{relative entropy}. This is appropriate and sometimes we will
follow this practice. However, since all entropies are relative, even when
relative to a uniform distribution, the qualifier `relative' is redundant
and can be dropped.

The functional $S[p,q]$ is designed by a process of elimination --- a
process of \emph{eliminative induction}. First we state the desired design
criteria; this is the crucial step that defines what makes one distribution
preferable over another. Then we analyze how each criterion constrains the
form of the entropy. As we shall see the design criteria adopted below are
sufficiently constraining that there is a single entropy functional $S[p,q]$
that survives the process of elimination.

This approach has a number of virtues. First, to the extent that the design
criteria are universally desirable, then the single surviving entropy
functional will be of universal applicability too. Second, the reason why
alternative entropy candidates are eliminated is quite explicit --- at least
one of the design criteria is violated. Thus, \emph{the justification behind
the single surviving entropy is not that it leads to demonstrably correct
inferences, but rather, that all other candidate entropies demonstrably fail
to perform as desired.}

\subsection{Specific design criteria}

\noindent \label{specific DC}Three criteria and their consequences for the
functional form of the entropy are given below. Proofs are given in \cite%
{Caticha2012}.

\subsubsection{Locality}

\begin{description}
\item[\textbf{DC1}] \emph{Local information has local effects.}
\end{description}

\noindent Suppose the information to be processed does \emph{not }refer to a
particular subdomain $\mathcal{D}$ of the space $\mathcal{X}$ of $x$s. In
the absence of any new information about $\mathcal{D}$ the PMU demands we do
not change our minds about probabilities that are conditional on $\mathcal{D}
$. Thus, we design the inference method so that $q(x|\mathcal{D})$, the
prior probability of $x$ conditional on $x\in \mathcal{D}$, is not updated.
The selected conditional posterior is\footnote{%
We denote priors by $q$, candidate posteriors by lower case $p$, and the
selected posterior by upper case $P$.} 
\begin{equation}
P(x|\mathcal{D})=q(x|\mathcal{D})~.  \label{DC1a}
\end{equation}%
We emphasize: the point is not that we make the unwarranted assumption that
keeping $q(x|\mathcal{D})$ unchanged is guaranteed to lead to correct
inferences. It need not; induction is risky. The point is, rather, that in
the absence of any evidence to the contrary there is no reason to change our
minds and the prior information takes precedence.

\noindent \textbf{The consequence of DC1} is that non-overlapping domains of 
$x$ contribute additively to the entropy, 
\begin{equation}
S[p,q]=\int dx\,F\left( p(x),q(x),x\right) \ ,  \label{DC1b}
\end{equation}%
where $F$ is some unknown function --- not a functional, just a regular
function of three arguments.

\noindent \textbf{Comment:}

If the variable $x$ is continuous the criterion DC1 requires that
information that refers to points infinitely close but just outside the
domain $\mathcal{D}$ will have no influence on probabilities conditional on $%
\mathcal{D}$. This may seem surprising as it may lead to updated probability
distributions that are discontinuous. Is this a problem? No.

In certain situations (\emph{e.g.}, physics) we might have explicit reasons
to believe that conditions of continuity or differentiability should be
imposed and this information might be given to us in a variety of ways. The
crucial point, however --- and this is a point that we keep and will keep
reiterating --- is that unless such information is in fact explicitly given
we should not assume it. If the new information leads to discontinuities, so
be it.

\noindent \textbf{Comment: Bayes' rule}

The locality criterion DC1 includes Bayesian conditionalization as a special
case. Indeed, if the information is given through the constraint $p(\mathcal{%
D})=1$ --- or more precisely $p(\mathcal{\tilde{D}})=0$ where $\mathcal{%
\tilde{D}}$ is the complement of $\mathcal{D}$ so that the information does
not directly refer to $\mathcal{D}$ --- then $P(x|\mathcal{D})=q(x|\mathcal{D%
})$, which is known as Bayesian conditionalization. More explicitly, if $%
\theta $ is the variable to be inferred on the basis of information about a
likelihood function $q(x|\theta )$ and observed data $x^{\prime }$, then the
update from the prior $q$ to the posterior $P$, 
\begin{equation}
q(x,\theta )=q(x)q(\theta |x)\rightarrow P(x,\theta )=P(x)P(\theta |x)
\end{equation}%
consists of updating $q(x)\rightarrow P(x)=\delta (x-x^{\prime })$ to agree
with the new data and invoking the PMU so that $P(\theta |x^{\prime
})=q(\theta |x^{\prime })$ remains unchanged. Therefore, 
\begin{equation}
P(x,\theta )=\delta (x-x^{\prime })q(\theta |x)~.
\end{equation}%
Marginalizing over $x$ gives 
\begin{equation}
P(\theta )=q(\theta |x^{\prime })=q(\theta )\frac{q(x^{\prime }|\theta )}{%
q(x^{\prime })}~,
\end{equation}%
which is Bayes' rule. Thus, \emph{entropic inference is designed to include
Bayesian inference as a special case}. Note however that imposing locality
is not identical to imposing Bayesian conditionalization --- locality is
more general because it is not restricted to absolute certainties such as $p(%
\mathcal{D})=1$.

\subsubsection{Coordinate invariance}

\begin{description}
\item[\textbf{DC2}] \emph{The system of coordinates carries no information. }
\end{description}

\noindent The points $x\in \mathcal{X}$ can be labeled using any of a
variety of coordinate systems. In certain situations we might have explicit
reasons to believe that a particular choice of coordinates should be
preferred over others and this information might have been given to us in a
variety of ways, but unless it was in fact given we should not assume it:
the ranking of probability distributions should not depend on the
coordinates used.

\noindent \textbf{The consequence of DC2} is that $S[p,q]$ can be written in
terms of coordinate invariants such as $dx\,m(x)$ and $p(x)/m(x)$, and $%
q(x)/m(x)$: 
\begin{equation}
S[p,q]=\int dx\,m(x)\Phi \left( \frac{p(x)}{m(x)},\frac{q(x)}{m(x)}\right) ~.
\label{DC2}
\end{equation}%
Thus the single unknown function $F$ which had three arguments has been
replaced by two unknown functions: $\Phi $ which has two arguments, and the
density $m(x)$.

To grasp the meaning of DC2 it may be useful to recall some facts about
coordinate transformations. Consider a change from old coordinates $x$ to
new coordinates $x^{\prime }$ such that $x=\Gamma (x^{\prime })$. The new
volume element $dx^{\prime }$ includes the corresponding Jacobian, 
\begin{equation}
dx=\gamma (x^{\prime })dx^{\prime }\quad \text{where}\quad \gamma (x^{\prime
})=\left\vert \frac{\partial x}{\partial x^{\prime }}\right\vert .
\label{coord jacobian}
\end{equation}%
Let $m(x)$ be any density; the transformed density $m^{\prime }(x^{\prime })$
is such that $m(x)dx=m^{\prime }(x^{\prime })dx^{\prime }$. This is true, in
particular, for probability densities such as $p(x)$ and $q(x)$, therefore 
\begin{equation}
m(x)=\frac{m^{\prime }(x^{\prime })}{\gamma (x^{\prime })}~,\quad p(x)=\frac{%
p^{\prime }(x^{\prime })}{\gamma (x^{\prime })}\quad \text{and}\quad q(x)=%
\frac{q^{\prime }(x^{\prime })}{\gamma (x^{\prime })}\,.
\label{coord trans dens}
\end{equation}%
The coordinate transformation gives

\begin{eqnarray}
S[p,q] &=&\int dx\,F\left( p(x),q(x),x\right)  \notag \\
&=&\int \gamma (x^{\prime })dx^{\prime }\,F\left( \frac{p^{\prime
}(x^{\prime })}{\gamma (x^{\prime })},\frac{q^{\prime }(x^{\prime })}{\gamma
(x^{\prime })},\Gamma (x^{\prime })\right) ,  \label{sa}
\end{eqnarray}%
which is a mere change of variables. The identity above is valid always, for
all $\Gamma $ and for all $F$; it imposes absolutely no constraints on $%
S[p,q]$. The real constraint arises from realizing that we could have \emph{%
started} in the $x^{\prime }$ coordinate frame, in which case we would have
ranked the distributions using the entropy 
\begin{equation}
S[p^{\prime },q^{\prime }]=\int dx^{\prime }\,F\left( p^{\prime }(x^{\prime
}),q^{\prime }(x^{\prime }),x^{\prime }\right) \,,  \label{sb}
\end{equation}%
but this should have no effect on our conclusions. This is the nontrivial
content of DC2. It is not that we can change variables, we can always do
that; but rather that the two rankings, the one according to $S[p,q]$ and
the other according to $S[p^{\prime },q^{\prime }]$ must coincide. This
requirement is satisfied if, for example, $S[p,q]$ and $S[p^{\prime
},q^{\prime }]$ turn out to be numerically equal, but this is not necessary.

\subsubsection{Locality (again)}

Next we determine the density $m(x)$ by invoking the locality criterion DC1
once again. A situation in which no new information is available is dealt by
allowing the domain $\mathcal{D}$ to cover the whole space of $x$s, $%
\mathcal{D=X}$ and DC1 requires that in the absence of any new information
the prior conditional probabilities should not be updated, $P(x|\mathcal{X}%
)=q(x|\mathcal{X})$ or $P(x)=q(x)$. Thus, when there are no constraints the
selected posterior distribution should coincide with the prior distribution,
which is expressed as

\begin{description}
\item[\textbf{DC1}$^{\prime }$] \emph{When there is no new information there
is no reason to change one's mind and one shouldn't.}
\end{description}

\noindent \textbf{The consequence of DC1}$^{\prime }$\textbf{\ }(a second
use of locality) is that the arbitrariness in the density $m(x)$ is removed:
up to normalization $m(x)$ must be the prior distribution $q(x)$, and
therefore at this point we have succeeded in restricting the entropy to
functionals of the form 
\begin{equation}
S[p,q]=\int dx\,q(x)\Phi \left( \frac{p(x)}{q(x)}\right) ~.  \label{DC1c}
\end{equation}

\subsubsection{Independence}

\begin{description}
\item[\textbf{DC3}] \emph{When two systems are a priori believed to be
independent and we receive independent information about each then it should
not matter whether one is included in the analysis of the other or not. }
\end{description}

\noindent Consider a composite system, $x=(x_{1},x_{2})\in \mathcal{X}=%
\mathcal{X}_{1}\times \mathcal{X}_{2}$. Assume that all prior evidence led
us to believe the systems were independent. This belief is expressed through
the prior distribution: if the individual priors are $q_{1}(x_{1})$ and $%
q_{2}(x_{2})$, then the prior for the whole system is $%
q_{1}(x_{1})q_{2}(x_{2})$. Further suppose that new information is acquired
such that $q_{1}(x_{1})$ would by itself be updated to $P_{1}(x_{1})$ and
that $q_{2}(x_{2})$ would be itself be updated to $P_{2}(x_{2})$. DC3
requires that $S[p,q]$ be such that the joint prior $%
q_{1}(x_{1})q_{2}(x_{2}) $ updates to the product $P_{1}(x_{1})P_{2}(x_{2})$
so that inferences about one system do not affect inferences about the other.

\noindent \textbf{The consequence of DC3 }is that the remaining unknown
function $\Phi $ is determined to be $\Phi (z)=-z\log z$. Thus, probability
distributions $p(x)$ should be ranked relative to the prior $q(x)$ according
to their relative entropy, 
\begin{equation}
S[p,q]=-\int dx\,p(x)\log \frac{p(x)}{q(x)}.  \label{S[p,q]}
\end{equation}

\noindent \textbf{Comment:}

We emphasize that the point is not that when we have no evidence for
correlations we draw the firm conclusion that the systems must necessarily
be independent. They could indeed have turned out to be correlated and then
our inferences would be wrong. Induction involves risk. The point is rather
that if the joint prior reflected independence and the new evidence is
silent on the matter of correlations, then the prior takes precedence and
there is no reason to change our minds. This is parsimony in action: a
feature of the probability distribution --- in this case, independence ---
will not be updated unless the evidence requires it.

\subsection{The ME\ method}

At this point our conclusions are summarized as follows:

\begin{description}
\item[\textbf{The ME method}: ] \emph{We want to update from a prior
distribution }$q$\emph{\ to a posterior distribution\ when there is new
information in the form of constraints }$\mathcal{C}$\emph{\ that specify a
family }$\{p\}$\emph{\ of allowed posteriors. The posterior is selected
through a ranking scheme that recognizes the value of prior information, the
irrelevance of choice of coordinates, and the privileged role of
independence. Within the family }$\{p\}$\emph{\ the preferred posterior }$P$ 
\emph{is that which maximizes the relative entropy }$S[p,q]$\emph{\ subject
to the available constraints. No interpretation for }$S[p,q]$\emph{\ is
given and none is needed.}
\end{description}

\noindent We emphasize that the logic behind the updating procedure does not
rely on any particular meaning assigned to the entropy, either in terms of
information, or heat, or disorder. Entropy is merely a tool for inductive
inference; we do not need to know what entropy means; we only need to know
how to use it.

The derivation above has singled out \emph{a unique }$S[p,q]$\emph{\ to be
used in inductive inference}. Other \textquotedblleft
entropies\textquotedblright\ such as those associated with the names of
Renyi or Tsallis might turn out to be useful for other purposes --- perhaps
as measures of some kinds of information, or measures of discrimination or
distinguishability among distributions, or of ecological diversity, or for
some altogether different function --- but they are unsatisfactory for the
purpose of updating in the sense that they do not perform according to the
design criteria DC1-3.

\subsection{Deviations from maximum entropy}

There is one last issue that must be addressed before one can claim that the
design of the method of entropic inference is more or less complete. Higher
entropy represents higher preference but there is nothing in the previous
arguments to tell us by how much. Suppose the maximum of the entropy
function is not particularly sharp, are we really confident that
distributions that are ranked close to the maximum are totally ruled out? We
want a quantitative measure of the extent to which distributions with lower
entropy are ruled out. The discussion below follows \cite{Caticha2000}.

The problem is to update from a prior $q(x)$ given information specified by
certain constraints $\mathcal{C}$. The constraints $\mathcal{C}$ specify a
family of candidate distributions $p(x)=p(x|\theta )$ which can be
conveniently labelled with some finite number of parameters $\theta ^{i}$, $%
i=1\ldots n$. Thus, the parameters $\theta $ are coordinates on a
statistical manifold $\Theta _{n}$ specified by $\mathcal{C}$. The
distributions in this manifold are ranked according to their entropy $%
S[p,q]=S(\theta )$ and the preferred posterior is the distribution $%
p(x|\theta _{0})$ that maximizes the entropy $S(\theta )$.

The question we now address concerns the extent to which $p(x|\theta _{0})$
should be preferred over other distributions with lower entropy or, to put
it differently: To what extent is it rational to believe that the selected
value ought to be the entropy maximum $\theta _{0}$ rather than any other
value $\theta $? This is a question about the probability $p(\theta )$ of
various values of $\theta $.

The original problem which led us to design the ME method was to assign a
probability to $x$; we now see that the full problem is to assign
probabilities to both $x$ and $\theta $.\ We are concerned not just with $%
p(x)$ but rather with the joint distribution $p_{J}(x,\theta )$; the
universe of discourse has been expanded from $\mathcal{X}$ (the space of $x$%
s) to the product space $\mathcal{X}\times \Theta _{n}$ (the space of $x$s
and $\theta $s).

To determine the joint distribution $p_{J}(x,\theta )$ we make use of
essentially the only method at our disposal --- the ME method itself --- but
this requires that we address the two standard preliminary questions: first,
what is the prior distribution, what do we know about $x$ and $\theta $
before we receive the information in the constraints $\mathcal{C}$? And
second, how do handle this new information $\mathcal{C}$ that constrains the
allowed $p_{J}(x,\theta )$?

This first question is the subtler one: when we know absolutely nothing
about the $\theta $s we do not know how they are related to the $x$s, and we
know neither the constraints $\mathcal{C}$ nor the space $\Theta _{n}$ they
determine. At best we just know that the $\theta $s are points in some
unspecified space $\Theta _{N}$ of sufficiently large dimension $N$. A joint
prior that reflects this state of ignorance is a product, $q_{J}(x,\theta
)=q(x)\mu (\theta )$. We will assume that the prior over $x$ is known --- it
is the same prior we had used when we updated from $q(x)$ to $p(x|\theta
_{0})$. We will also assume that $\mu (\theta )$ represents a uniform
distribution, that is, it assigns equal probabilities to equal volumes in $%
\Theta _{N}$.

Next we tackle the second question: what are the constraints on the allowed
joint distributions $p_{J}(x,\theta )=p(\theta )p(x|\theta )$? The new
information is that the $\theta $s are not any arbitrary points in some
unspecified large space $\Theta _{N}$ but are instead constrained to lie on
the smaller subspace $\Theta _{n}$ that represents those distributions $%
p(x|\theta )$ satisfying the constraints $\mathcal{C}$. This space $\Theta
_{n}$ is a statistical manifold and there exists a natural measure of
distance given by the information metric $g_{ij}$. The corresponding volume
elements are given by $g_{n}^{1/2}(\theta )d^{n}\theta $, where $%
g_{n}(\theta )$ is the determinant of the metric \cite{Caticha2012} \cite%
{Amari1985}. Therefore, on the constraint manifold $\Theta _{n}$ the uniform
prior for $\theta $ is proportional to $g_{n}^{1/2}(\theta )$ and the
corresponding joint prior is $q_{J}(x,\theta )=q(x)g_{n}^{1/2}(\theta )$.

To select the preferred joint distribution $P(x,\theta )$ we maximize the
joint entropy $\mathcal{S}[p_{J},q_{J}]$ over all distributions of the form $%
p_{J}(x,\theta )=p(\theta )p(x|\theta )$ with $\theta \in \Theta _{n}$. It
is convenient to write the joint entropy as 
\begin{eqnarray}
\mathcal{S}[p_{J},q_{J}] &=&-\int_{\mathcal{X}\times \Theta
_{n}}dx\,d^{n}\theta \,p(\theta )p(x|\theta )\,\log \frac{p(\theta
)p(x|\theta )}{g_{n}^{1/2}(\theta )q(x)}  \notag \\
&=&-\int_{\Theta _{n}}\,d^{n}\theta \,p(\theta )\log \frac{p(\theta )}{%
g_{n}^{1/2}(\theta )}+\int_{\Theta _{n}}d^{n}\theta \,p(\theta )S(\theta ),
\label{S[joint]}
\end{eqnarray}%
where 
\begin{equation}
S[p,q]=S(\theta )=-\int_{\mathcal{X}}\,dx\,p(x|\theta )\log \frac{p(x|\theta
)}{q(x)}.  \label{Stheta}
\end{equation}%
Then, maximizing (\ref{S[joint]}) with respect to variations $\delta
p(\theta )$ such that $\int d^{n}\theta \,p(\theta )=1$, yields 
\begin{equation}
0=\int_{\Theta _{n}}\,d^{n}\theta \left( -\log \frac{p(\theta )}{%
g^{1/2}(\theta )}+S(\theta )+\log \zeta \right) \,\delta p(\theta )\,.
\end{equation}%
(The required Lagrange multiplier has been written as $1-\log \zeta $.)
Therefore the probability that the value of $\theta $ should lie within the
small volume $dV_{n}=g_{n}^{1/2}(\theta )d^{n}\theta $ is 
\begin{equation}
P(\theta )d^{n}\theta =\frac{1}{\zeta }\,\,e^{S(\theta )}dV_{n}\quad \text{%
with\quad }\zeta =\int_{\Theta _{n}}dV_{n}\,e^{S(\theta )}.  \label{main}
\end{equation}%
Equation (\ref{main}) is the result we seek. It tells us that, as expected,
the preferred value of $\theta $ is the value $\theta _{0}$ that maximizes
the entropy $S(\theta )$, eq.(\ref{Stheta}), because this maximizes the 
\emph{scalar} probability density $\exp S(\theta )$.\footnote{%
The density $\exp S(\theta )$ is a scalar function; it is the probability
per unit invariant volume $dV=g_{n}^{1/2}(\theta )d^{n}\theta $.} It also
tells us the degree to which values of $\theta $ away from the maximum $%
\theta _{0}$ are ruled out.

The previous discussion allows us to refine our understanding of the ME
method. ME\ is not an all-or-nothing recommendation to pick that single
distribution that maximizes entropy and rejects all others. The ME\ method
is more nuanced: in principle all distributions within the constraint
manifold ought to be included in the analysis; they contribute in proportion
to the exponential of their entropy and this turns out to be significant in
situations where the entropy maximum is not particularly sharp.

Going back to the original problem of updating from the prior $q(x)$ given
information that specifies the manifold $\{p(x|\theta )\}$, the preferred
update within the family $\{p(x|\theta )\}$ is $p(x|\theta _{0})$, but to
the extent that other values of $\theta $ are not totally ruled out, a
better update is obtained marginalizing the joint posterior $P_{J}(x,\theta
)=P(\theta )p(x|\theta )$ over $\theta $, 
\begin{equation}
P_{\text{ME}}(x)=\int_{\Theta _{n}}d^{n}\theta \,P(\theta )p(x|\theta
)=\int_{\Theta _{n}}dV_{n}\frac{\,e^{S(\theta )}}{\zeta }p(x|\theta )~.
\label{ME posterior}
\end{equation}%
In situations where the entropy maximum at $\theta _{0}$ is very sharp we
recover the old result, 
\begin{equation}
P_{\text{ME}}(x)\approx p(x|\theta _{0})~.
\end{equation}%
When the entropy maximum is not very sharp eq.(\ref{ME posterior}) is the
more honest update.

The summary description of the ME method in the previous subsection can now
be refined by adding the following line:

\begin{description}
\item[\qquad ] \emph{The ME posterior }$P_{\text{ME}}$\emph{\ is a weighted
average of all distributions in the family }$\{p\}$\emph{\ specified by the
constraints }$\mathcal{C}$\emph{. Each }$p$\emph{\ is weighted by the
exponential of its entropy }$S[p,q]$\emph{. }
\end{description}

Physical applications of the extended ME method are ubiquitous. For
macroscopic systems the preference for the distribution that maximizes $%
S[p,q]$ can be overwhelming but for small systems such fluctuations about
the maximum are common. Thus, violations of the second law of thermodynamics
can be seen everywhere --- provided we know where to look. For example, eq.(%
\ref{main}) agrees with Einstein's theory of thermodynamic fluctuations and
extends it beyond the regime of small fluctuations. Another important
application, developed in \cite{Caticha2011}, is quantum mechanics --- the
ultimate theory of small systems.

\section{Summary}

Science requires a framework for inference on the basis of incomplete
information. We showed how to design tools to represent a state of partial
knowledge as a web of interconnected beliefs with no internal
inconsistencies; the resulting scheme is probability theory. Then we argued
that in a properly Bayesian framework the concept of information must be
defined in terms of its effects on the beliefs of rational agents. The
definition we have proposed -- that information is a constraint on rational
beliefs -- is convenient for two reasons. First, the information/belief
relation is explicit, and second, such information is ideally suited for
quantitative manipulation using entropic\ methods. Finally, we designed a
method for updating probabilities. The design criteria are strictly
pragmatic; the method aims to be of universal applicability, it recognizes
the value of information both old and new, and it recognizes the special
status that must be accorded to considerations of independence in order to
build models that are actually usable. The result --- the maximum entropy or
ME\ method --- is a framework in which entropy is the tool for updating
probabilities. The ME method unifies both MaxEnt and Bayes' rule into a
single framework of inductive inference and allows new applications. Indeed,
much as the old MaxEnt method provided the foundation for statistical
mechanics, recent work has shown that the extended ME method provides an
entropic foundation for quantum mechanics.

Within an informational approach it is not possible to sharply separate the
subject matter or contents of science from the inductive methods of science;
science includes both. This point of view has two important consequences.
The first is that just as we accept that the contents of science will evolve
over time, to the extent that contents and methods are not separable, we
must also accept that the inference methods are provisional too --- the best
we currently have --- and are therefore susceptible to future change and
improvement.

The second consequence stems from the observation that experiments do not
vindicate individual propositions within a theory; they vindicate the theory
as a whole. Therefore, when a theory turns out to be pragmatically
successful we can say that it is not just the contents of the theory that
have been corroborated but also its methods of inference. Thus, the ultimate
justification of the entropic methods of inference resides in their
pragmatic success when confronted with experiment.

\section{Discussion}

The accidents of history have caused the term `pragmatism' to acquire
connotations that are not fully satisfactory. But we do not have to
literally adopt Peirce's version of pragmatism or James' or Putnam's. It is
not necessary that we agree with everything or even most of what these
authors have said; not only did they not agree with each other, but their
views evolved, and their later views disagreed with those held in their own
youth. Instead, the proper pragmatic attitude is to pick and choose useful
bits and pieces and try to stitch them into a coherent framework. To the
extent that this framework turns out to be useful we have succeeded and that
is all we need.

The vague notion of truth that I have favored --- truth as a compliment ---
is not Peirce's. It might be closer to James' or Putnam's and even
Einstein's --- but it need not be identical to their notions either: the
real purpose is not a theory of truth itself but rather those applications
that may be tackled through a pragmatically designed framework for
inference. In other words, the theory of truth is not the real goal; it is
only an intermediate obstacle, perhaps even a distraction, on the way to the
real problems: Can we do quantum mechanics? Can we do economics? Or,
borrowing Floridi's words, can we \textquotedblleft model the world in such
a way to make sense of it and withstand its impact\textquotedblright ?

I\ have argued in favor of an informational pragmatic realism but I have not
attempted its systematic development. Some of its features can be summarized
as follows: There is a world out there which we must navigate. Physical
models are inference schemes; they are instruments to help us succeed.
Within such models we find elements that purport to represent entities such
as particles or fields. We also find other elements that are tools for
manipulating information --- probabilities, entropies, wave functions. All
models inevitably involve concepts and categories of our own construction
chosen for our own pragmatic reasons. To the extent that a model is reliably
successful we say that its entities are real. Beyond that it is meaningless
to assert that these entities enjoy any special relation to the world that
might be described as \textquotedblleft referring\textquotedblright\ or
\textquotedblleft corresponding\textquotedblright\ to something independent
of us.

Within the informational approach to science the notion of a structural
realism in which the world consists of relations and structures only ---
without any entities or objects --- makes no sense. The information or lack
thereof is information about something --- both the entities and the
informational tools must appear in the models. Having said that, the central
point of the informational approach to physics \cite{Jaynes1957} \cite%
{Caticha2011} \cite{Caticha2012} is precisely that the formal rules for
manipulating information --- Bayesian and entropic methods --- place such
strong constraints on the formal structure of theories that a label of \emph{%
informational structural pragmatic realism} may, in the end, be quite
appropriate.

Therefore I welcome Floridi's reconciliation of the epistemic and the
(non-eliminativist) ontic versions of structural realism. Indeed, the
recognition that a model describes a structure at a given level of
abstraction brings Floridi close to Putnam's internal realism. `Internal'
because the entities and structures in our world are defined \emph{within} a
given level of abstraction or conceptual framework which is chosen by us for
our purposes. And `realism' because the conceptual framework is not
arbitrary; it is not independent of the world; the chosen entities and
structures have to be useful and succeed in the real world. An important
difference, however, is that Putnam's internal realism makes no mention of
information. This is the gap that can hopefully be closed by the
informational pragmatic realism advocated here.

Another aspect of Floridi's structural realism that I find appealing is the
idea that the distinction between what constitutes intrinsic nature and what
constitutes structure is not easily drawn --- that relata are not logically
prior to relations, that they come together, all or nothing, in a single
package. Such an idea is already deeply ingrained in physics. For example, a
quantity such as electric charge refers on one hand to an intrinsic property
of a particle --- an electron without its charge would just not be an
electron --- and on the other hand electric charge describes interactions,
the relations of the particle to other charged particles. It is totally
inconceivable to claim that a particle could possible have an electric
charge and yet not interact with other particles according to the laws of
electromagnetism. Therefore, it is perfectly legitimate to assert that an
electron is, to use Floridi's term, a structural object --- its intrinsic
properties are defined by the structure of its relations to other particles.
Furthermore, as shown in \cite{Caticha2011} the electric interactions can
indeed be approached from a purely informational perspective. They are
described through information --- that is, constraints --- on the allowed
motions that the particle can undertake.

The definition of information that I have proposed --- information as a
constraint on rational beliefs --- differs from Floridi's definition as
well-formed, meaningful, and truthful data. But there is considerable
overlap. Indeed, within the entropic/Bayesian framework a mere set of
numbers --- or data --- does not by itself constitute information. It is
necessary that the data be embedded within a model. This is what endows the
data with significance, with meaning. The model provides a connection
between the data and the other quantities one wishes to infer --- a relation
that is usually established through what in statistics is called a
likelihood function. Moreover, as discussed in section 5, the data becomes
\textquotedblleft information\textquotedblright\ precisely at that moment
when we feel justified in allowing its effects to propagate throughout the
web of beliefs. Such data deserve the compliment `true'. And thus data is
information provided it is well-formed and meaningful by virtue of being
embedded in a model, and is truthful by virtue of an appropriate judgement.

Pragmatism is a form of empiricism too, at least in the sense that,

\begin{description}
\item[\qquad ] \textquotedblleft ...it is contented to regard its most
assured conclusions concerning matters of fact as hypotheses liable to
modification in the course of future experience...\textquotedblright\ (\cite%
{James1897}, p.vii)
\end{description}

\noindent van Fraasen's empiricism, either in its earlier form of \emph{%
constructive empiricism} or in the later form of \emph{empiricist
structuralism}, appears to adopt a correspondence model of truth and this
places him squarely in the anti-realist camp. Nevertheless, if we ignore the
labels and just look at what he actually wrote we find significant points of
contact with both structural and pragmatic realism. van Fraasen has argued
that in a semantic approach to theories as mathematical models if one
mathematical structure can represent the phenomena then any other isomorphic
structure can also do it:

\begin{description}
\item[\qquad ] \textquotedblleft ...models represent nature only up to
isomorphism -- they only represent structure.\textquotedblright\ \cite%
{vanFraasen1997}
\end{description}

\noindent This has been called the underdetermination \emph{problem} but we
will, more optimistically, regard it as an \emph{opportunity}. It offers the
possibility of imposing additional pragmatic virtues such as explanatory
power, simplicity, etc., that go beyond mere empirical adequacy, as criteria
for the acceptance of theories. This opportunity has been advantageously
pursued by Ellis \cite{Ellis1985}, and Floridi (\cite{Floridi2011},
p.358-360). van Fraasen too recognizes that the acceptance of a theory rests
on pragmatic considerations that go beyond mere belief in empirical
adequacy. (\cite{vanFraasen1980}, p. 12-13).

From our pragmatic perspective two of van Fraasen's ideas appear
particularly appealing. One has to do with clarifying the meaning of
empirical adequacy \cite{vanFraasen1997} \cite{vanFraasen2006b}. The problem
is that under a correspondence model of truth one might naively attempt to
achieve an agreement between the model and the phenomena as they are in
themselves --- which leads to the same old problem of reference. van Fraasen
skillfully evades this problem by asserting that empirical adequacy is not
adequacy to the phenomena-in-themselves but rather to the \emph{phenomena as
described by us}. This is as pragmatic as it gets! Empirical adequacy
involves the comparison of two descriptions; one is supplied by the theory
and the other is our description --- a \textquotedblleft data
model\textquotedblright\ --- of certain selectively chosen aspects that are
relevant to our interests, a description in terms of concepts invented by us
because they are adequate to our purposes. van Fraasen sums up as follows:

\begin{description}
\item[\qquad ] \textquotedblleft ... in a context in which a given model is
my representation of a phenomenon, there is no difference between the
question of whether a theory fits that representation and the question of
whether it fits the phenomenon.\textquotedblright\ \cite{vanFraasen2006b}
\end{description}

\noindent It is difficult to imagine that either James or Putnam would have
objected.

The second appealing idea concerns the transition from an older theory to a
newer theory that is presumably better --- van Fraasen calls it the issue of
royal succession in science \cite{vanFraasen2006a}. This topic was briefly
mentioned in section 1 with reference to the pessimistic meta-induction and
the no-miracles arguments.

Any systematic procedure for theory revision would naturally attempt to
preserve those features of the old theory that made it work. Which leads to
the question \textquotedblleft what makes a good theory
good?\textquotedblright\ Whatever it is --- it could perhaps be that it
captures the correct structure --- we can call it \textquotedblleft
true\textquotedblright\ or \textquotedblleft real\textquotedblright\ but
this is just a name, a compliment that merely indicates success. Explaining
the predictive success of science by \textquotedblleft
realism\textquotedblright\ or by \textquotedblleft truth\textquotedblright\
is somewhat analogous to explaining that opium will put you to sleep because
it has \textquotedblleft dormitive\textquotedblright\ powers.

Instead I find van Fraasen's pragmatic argument much more persuasive. The
success of science is not a miracle --- it is a matter of how scientific
theories are designed: beyond retaining the empirical successes of the old
theory we want more and the new theory must provide us with new empirical
successes. This is a necessary requirement for the new theory to be
accepted. Thus today's science is bound to be more successful than
yesterday's --- otherwise we would not have made the switch.

Within the pragmatic framework advocated here royal succession is explained
by adopting a relaxed form of van Fraasen's criteria. The acceptance
criteria are extended beyond mere empirical success to include other
pragmatic virtues: the new theory must retain the pragmatic successes of the
old and add a few of its own. The point is that it is not strictly necessary
that the new theory must lead to new empirical successes; it might, for
example, just lead to better explanations, or be more computationally
convenient.

I conclude that an \emph{informational pragmatic realism} has the potential
of incorporating many valuable pragmatic insights derived from internal
realism, informational structural realism and empiricist structuralism into
a single coherent doctrine and thereby close the gap between them.

\noindent \textbf{Acknowledgements: }I am grateful to C. Cafaro, N. Caticha,
A. Giffin, A. Golan, P. Goyal, K. Knuth, C. Rodr\'{\i}guez, M. Reginatto,
and J. Skilling for many useful discussions on entropic inference; and also
to A. Beavers and L. Floridi for the invitation to participate in this
symposium.

\end{document}